\documentclass[aps,showpacs,preprint,pre,superscriptaddress]{revtex4}
\usepackage{graphicx}
\input amssym.tex

\begin{document}
\title{Comment on ''Exceptional points and double poles of the $S$ matrix''}
\author{C.~Dembowski}
\affiliation{Institut f{\"u}r Kernphysik, Technische Universit{\"a}t
Darmstadt, D-64289 Darmstadt, Germany}
\author{B.~Dietz}
\affiliation{Institut f{\"u}r Kernphysik, Technische Universit{\"a}t
Darmstadt, D-64289 Darmstadt, Germany}
\author{H.-D.~Gr{\"a}f}
\affiliation{Institut f{\"u}r Kernphysik, Technische Universit{\"a}t
Darmstadt, D-64289 Darmstadt, Germany}
\email{hanns-ludwig.harney@mpi-hd.mpg.de} \author{H.L.~Harney}
\affiliation{Max-Planck-Institut f{\"u}r Kernphysik, D-69029
Heidelberg, Germany}
\author{A.~Heine}
\affiliation{Institut f{\"u}r Kernphysik, Technische Universit{\"a}t
Darmstadt, D-64289 Darmstadt, Germany}
\author{W.~D.~Heiss}
\affiliation{Department of Physics, University of Stellenbosch, 7602
Matieland, South Africa}
\author{A.~Richter}
\affiliation{Institut f{\"u}r
Kernphysik, Technische Universit{\"a}t Darmstadt, D-64289 Darmstadt,
Germany}
\date{\today}

\begin{abstract}
In a recent paper [Phys.~Rev.~E \textbf{67}, 026204 (2003)], Rotter
calculates the geometric phases that are picked up by the eigenvectors
of a two-state quantum system when an exceptional point is encircled.
In addition to the geometric phases observed by Dembowski \emph{et al.}
in a microwave cavity experiment [Phys.~Rev.~Lett.~\textbf{86}, 787
(2001)], she finds a phase factor of $i$. We show that Rotters results
are inconsistent.
\end{abstract}
\pacs{05.45.Mt,03.65.Vf}
\maketitle

In \cite{Dembo:01} an exceptional point (EP), i.e. a branch point
singularity of fourth order \cite{Heiss:99}, of a parametric
dissipative microwave cavity has been observed experimentally by
tracking the development of the eigenvalues and eigenvectors of the
resonator while the EP is encircled. It has been found that along the
closed loop around the EP the eigenvectors pick up geometrical phases
in accordance with \cite{Heiss:99}.

In a recent paper \cite{Rotter:03}, Rotter criticizes this experiment
and suggests the existence of an unobserved additional geometrical
phase factor of $i$. This suggestion is inconsistent.

In \cite{Rotter:03}, Rotter states in Eq.~(24) her transformation
scheme for the two eigenvectors, $\tilde{\Phi}_1$ and $\tilde{\Phi}_2$,
when an EP is encircled (denoted in \cite{Rotter:03} by the symbol
'$\rightarrow$'):
\begin{equation}
\left\{\tilde{\Phi}_1,\tilde{\Phi}_2\right\}\rightarrow
\left\{-i\tilde{\Phi}_2,+i\tilde{\Phi}_1\right\}
\label{EQRotter1R}
\end{equation}
A second full surrounding with the same orientation gives according to
Eq.~(25) of \cite{Rotter:03}:
\begin{equation}
\left\{-i\tilde{\Phi}_2,+i\tilde{\Phi}_1\right\}\rightarrow
\left\{+\tilde{\Phi}_1,+\tilde{\Phi}_2\right\}
\label{EQRotter2R}
\end{equation}
By reversing this second loop, i.e. reading Eq.~(\ref{EQRotter2R}) from
right to left, the transformation scheme for encircling an EP
\emph{with the opposite orientation} directly follows to be:
\begin{equation}
\left\{\tilde{\Phi}_1,\tilde{\Phi}_2\right\}\rightarrow
\left\{-i\tilde{\Phi}_2,+i\tilde{\Phi}_1\right\}
\label{EQRotter1La}
\end{equation}
In Eq.~(27) of \cite{Rotter:03}, Rotter states however, that encircling
the EP with the orientation opposite to that of Eq.~(\ref{EQRotter2R})
yields:
\begin{equation}
\left\{\tilde{\Phi}_1,\tilde{\Phi}_2\right\}\rightarrow
\left\{+i\tilde{\Phi}_2,-i\tilde{\Phi}_1\right\}
\label{EQRotter1Lb}
\end{equation}
This contradicts her earlier result: Equations~(\ref{EQRotter1La}) and
(\ref{EQRotter1Lb}) are incompatible with each other.

Since Rotters statements are inconsistent we conclude that the
additional phase factor of $i$ as suggested in \cite{Rotter:03} and
other articles of her \cite{RotterVarious} is not present.


\begin{thebibliography}{99}

\bibitem{Dembo:01}C.~Dembowski, H.-D.~Gr\"af, H.L.~Harney, A.~Heine,
W.D.~Heiss, H.~Rehfeld and A.~Richter, Phys.~Rev.~Lett.~\textbf{86},
787 (2001).

\bibitem{Heiss:99} W.D.~Heiss, M.~M{\"u}ller, and I.~Rotter,
Phys.~Rev.~E~\textbf{58}, 2894 (1998); W.D.~Heiss, Eur.~Phys.~J.~D
\textbf{7}, 1 (1999); Phys.~Rev.~E~\textbf{61}, 929 (2000).

\bibitem{Rotter:03} I.~Rotter, Phys.~Rev.~E \textbf{67}, 026204 (2003).

\bibitem{RotterVarious}I.~Rotter, Phys.~Rev.~E~\textbf{65}, 026217
(2002) and J.~Okolowisc, M.~Ploszajczak and I.~Rotter,
Phys.~Rep.~\textbf{374}, 271 (2003).

\end{thebibliography}
\end{document}